\documentclass[oribibl]{llncs}

\usepackage{amssymb}
\usepackage{amsmath}

\usepackage[colorlinks=true, linkcolor=blue, citecolor=blue, urlcolor=blue]{hyperref}

\usepackage{graphicx}
\usepackage{booktabs}
\usepackage{multirow}
\usepackage{tabularx}
\usepackage{enumitem}
\usepackage{threeparttable}
\usepackage{balance}

\usepackage{url}

\DeclareMathOperator{\WebPage}{\mathcal{P}}
\DeclareMathOperator{\JavaScript}{\mathcal{J}}
\DeclareMathOperator{\NA}{\varnothing}
\DeclareMathOperator{\URL}{\mathcal{U}}
\DeclareMathOperator{\hash}{\mathcal{H}}
\DeclareMathOperator{\continuous}{\mathcal{C}}
\DeclareMathOperator{\dichotomous}{\mathcal{D}}
\newcommand{\hashset}{H_{\URL}}
\DeclareMathOperator{\SOP}%
{\stackrel{\text{\tiny SOP}}{\equiv}}
\DeclareMathOperator{\CO}%
{\stackrel{\text{\tiny SOP}}{\not\equiv}}

\usepackage{fancyhdr} % for arxiv
\fancypagestyle{firststyle}
{
   \fancyhf{}
   \fancyfoot[C]{\scriptsize{Proceedings of the 33rd IFIP International Conference on ICT Systems Security and Privacy Protection (IFIP SEC 2018), Pozn\'an, Springer, pp.~385--398. The final publication is available at Springer via \url{https://doi.org/10.1007/978-3-319-99828-2_27}}}
}

\begin{document}

\title{On the Integrity of Cross-Origin JavaScripts}

% for arxiv
\author{Jukka Ruohonen \and Joonas Salovaara \and Ville Lepp\"anen \\ \email{\{juanruo, joosal, ville.leppanen\}@utu.fi}}
%\author{Jukka Ruohonen\orcidID{0000-0001-5147-3084} \and Joonas Salovaara \and Ville Lepp\"anen \\ \email{\{juanruo, joosal, ville.leppanen\}@utu.fi}}
\institute{Department of Future Technologies, University of Turku, Finland}

\maketitle

\begin{abstract}
The same-origin policy is a fundamental part of the Web. Despite the restrictions imposed by the policy, embedding of third-party JavaScript code is allowed and commonly used. Nothing is guaranteed about the integrity of such code. To tackle this deficiency, solutions such as the subresource integrity standard have been recently introduced. Given this background, this paper presents the first empirical study on the temporal integrity of cross-origin JavaScript code.  According to the empirical results based on a ten day polling period of over $35$ thousand scripts collected from popular websites, (i) temporal integrity changes are relatively common; (ii) the adoption of the subresource integrity standard is still in its infancy; and (iii) it is possible to statistically predict whether a temporal integrity change is likely to occur. With these results and the accompanying discussion, the paper contributes to the ongoing attempts to better understand security and privacy in the current Web.
\end{abstract}


\section{Introduction}

\thispagestyle{firststyle} % for arxiv

Most current websites load numerous resources from many distinct third-party sources. Among these resources is JavaScript that is executed by clients visiting the websites. There are many viewpoints to this execution of third-party code.

One relates to network protocols \cite{Kumar17}. The \textit{hypertext transfer protocol} (HTTP) over the transport layer security protocol (a.k.a.~HTTPS) can only authenticate the server to which a client connects. It does not provide any guarantees about the authenticity of the encrypted content transmitted after the authentication. From this perspective, the authenticity (integrity) of web content has become a pressing concern as more and more content is transmitted through \textit{content delivery networks} (CDNs) and cloud services, while at the same time legislations all over the world have seen amendments toward mass surveillance. Another viewpoint relates to privacy \cite{Nikiforakis12, SomeBielova17}. In many respects, the execution of arbitrary third-party code is in the interests of those involved in the tracking of the Web's client-side. A further viewpoint relates to web security and web standards. 

The execution of third-party JavaScript occurs in the same context as the execution of the primary code present in a website \cite{Bielova13, Magazinius14}. To patch this limitation, the so-called \textit{subresource integrity} standard has recently been introduced for allowing enumeration of cryptographic hashes that clients verify before execution.\footnote{~In this paper, the term \textit{standard} includes also recommendations, guidelines, and working drafts that are well-recognized but not necessarily yet officially standardized.} Although the standard is oddly restricted only to certain web elements \cite{Cucurull16}, it is an important step toward at least some theoretical integrity guarantees. The standard and  associated considerations are also adopted as frames for this paper. In particular, the paper's motivation builds on the practical challenges facing the widespread adoption of the standard. As will be elaborated, it is also these practical challenges through which the wider security and privacy viewpoints can be reflected. To these ends, the following contributions are made:

\begin{enumerate}
\item{The paper presents the first empirical study on the temporal (data) integrity of cross-origin JavaScript code used and executed in the wild.}
\item{The paper shows that temporal integrity changes are relatively common on one hand and subresource integrity checks very uncommon on the other.}
\item{The paper demonstrates that a limited set of information can be used to predict whether cross-origin JavaScript code is likely to change temporally.}
\end{enumerate}

The remainder of the paper is structured into four straightforward sections. Namely: Section~\ref{section: background} discusses the background, Section~\ref{section: data} introduces the dataset, Section~\ref{section: results} presents the results, and Section~\ref{section: discussion} concludes with a few remarks.

\section{Background}\label{section: background}

In what follows, the rationale for the empirical study is motivated by briefly discussing the background related to remote cross-origin JavaScript inclusions.

\subsection{The Same-Origin Policy}

The \textit{same-origin policy} (SOP) is a fundamental part of the Web. It governs the ways elements in a \textit{hypertext markup language} (HTML) document can interact. An origin is defined as a tuple containing a scheme, a host name, and a port~\cite{RFC6454}. The tuple can be elaborated with the syntax for uniform resource identifiers:
\begin{equation}\label{eq: uri}
\texttt{scheme://}\underbrace{[\texttt{user}:\texttt{password}@]\texttt{host}:\texttt{port}}_{\texttt{authority}}/\texttt{path?query\#fragment} ,
\end{equation} 
where \texttt{host} is a fully qualified domain name or an Internet protocol address~\cite{RFC2396}. If and only if the \texttt{scheme}, the \texttt{host}, and the \texttt{port} fields are equal between two \textit{uniform resource locators} (URLs), the two locators have the same origin. When this condition is not satisfied, the antonym term \textit{cross-origin} is often used. 

A couple of additional points are warranted about the syntax and its semantics. The first  point is about the \texttt{scheme}. The inclusion of this protocol field is essential for isolating websites served via plain HTTP from those served via HTTPS~\cite{RFC6454}. The second point is about the \texttt{port} field: when it is missing, the information is derived from the mandatory \texttt{scheme}. Thus, the same-origin condition holds for the two tuples within the following two example URLs:
\begin{equation}\label{eq: url}
\texttt{http://example.com/index.html}
\stackrel{\text{\tiny SOP}}{\equiv}
\texttt{http://example.com:port} 
\end{equation}
when either $\texttt{port} \equiv 80$ or the two URLs are queried with Internet Explorer, which disregards \texttt{port} when deducing about origins \cite{Mozilla18b}. For this particular web browser, the tuples from \eqref{eq: url} have the same origin for any $\texttt{port} \in [1, 65535]$. 

% Re-read Bienlova13.
%
The SOP is used for many functions explicitly or implicitly related to privilege separation~\cite{Bielova13, DongHu13}. While these functions cover numerous web elements, the most important function is to restrict the execution of JavaScript by a web browser (refer to \cite{Zalewski09} for slightly outdated but still useful, extensive technical discussion). In essence, two same-origin documents have full access to each other's web resources. They can make HTTP requests to each other via JavaScript, they can manipulate each other's \textit{document object model} (DOM) that acts as an interface between JavaScript and HTML, and they can even share information about cookies. Thus, without the SOP, a JavaScript running in one tab of a user's browser could do practically anything with the content in another tab. 

% For good JSONP discussion read:
% https://jvaneyck.wordpress.com/2014/01/07/cross-domain-requests-in-javascript/

Despite the SOP restrictions, \textit{cross-origin embedding} is often allowed~\cite{Mozilla18b}. In particular, cross-origin requests are allowed for \texttt{<script>} tags equipped with the \texttt{src} attribute. When a client's browser encounters such a tag, it will issue a HTTP GET request to retrieve the content, which is executed immediately in the current origin's context. Although additional (security) information can be added to the GET requests via the \texttt{query} and \texttt{fragment} fields~\cite{Jayaraman10}, the basic security issue thus is that the JavaScript response has full privileges within the requesting web page \cite{DeRyck10, Magazinius14}. In addition, the \textit{cross-origin resource sharing} (CORS) standard \cite{W3C14a} can be used to relax the SOP policy by whitelisting (in HTTP header responses) those hosts from which cross-origin resources can be loaded dynamically with JavaScript. Either way, the fundamental security risk remains identical: if the host behind a script's source is compromised, arbitrary code can be executed in the context of all websites having included the script. The risk is not only theoretical; a recent data breach allegedly involved a single line of misplaced HTML and a compromised third-party~\cite{ITWire08}. One way to analytically approach the risk is to consider the integrity of remote JavaScripts.

\subsection{Integrity of Cross-Origin JavaScripts}\label{subsec: integrity of cross-origin javascripts}

There are numerous distinct aspects to the integrity of websites. For instance, at the DOM and HTML levels, integrity ``ensures that the contents of an interaction cannot be modified without the knowledge of the interacting components"~\cite{DeRyck10}. Another example would be the concept of web session integrity, which ``ensures that an attacker can never force the browser into introducing unintended messages in sessions established with trusted websites, or into leaking the authentication credentials (cookies and passwords) associated to these sessions" \cite{Bugliesi17}. When remote JavaScripts are included in a website, a prerequisite for this notion of web session integrity is the integrity of the remote JavaScripts themselves. For this purpose, the traditional notion of data integrity is suitable.

Thus, let $h(\cdot)$ denote a cryptographic hash function and $\hash$ a message digest outputted by the function for a given input. (For Internet measurement research, the first secure hash algorithm, or SHA-1, is sufficient and used also in the empirical part of this paper.) Let $\JavaScript$ further denote a unique cross-origin JavaScript that was included by a web page $\WebPage$ with a \texttt{<script>} tag whose \texttt{src} attribute pointed to a unique uniform resource locator, $\URL$. Consider then that this  URL,
\begin{equation}\label{eq: cross-origin URLs}
\URL \in \WebPage 
\quad\textmd{but}\quad
\URL \CO \WebPage ,
\end{equation}
was downloaded at time index $t$ through
%s
\begin{equation}\label{eq: download}
f(\URL, t) =
\begin{cases}
\JavaScript_{t} & \textmd{if HTTP GET of}~\URL~\textmd{was succesful at time}~t~\textmd{and} \\
\NA & \textmd{otherwise (including all network errors, etc.)}
.
\end{cases}
\end{equation}

By further defining $t = 1, 2, \ldots, T$ and 
\begin{equation}
h(f(\URL, t)) = 
\hash_{t}~\textmd{if}~f(\URL, t) \not\equiv \NA
~\textmd{and}~\NA~\textmd{otherwise} ,
\end{equation}
the integrity of the remote $\JavaScript$ from the viewpoint of $\WebPage$ can be evaluated by comparing the output from $h(\cdot)$ for two inputs $f(\URL, t)\not\equiv\NA$ and $f(\URL, t + k)\not\equiv\NA$ with $k \neq 0$ and $0 < t + k \leq T$. To a reasonable degree for empirical research purposes, the integrity has been intact if and only if $\hash_t \equiv \hash_{t+k}$ (see \cite{Geihs16} for a formal exposition of the same temporal idea). In theory, an integrity violation might be defined to occur also when  $h(f(\URL, t)) \equiv \hash_t$ but $h(f(\URL, t + k)) \equiv \NA$. As the Internet is not entirely reliable in terms of content transmission via either HTTP or HTTPS, the weaker definition is used in the empirical analysis.

The use of cryptographic hashes has been adopted also for recent web standards, recommendations, and guidelines. Namely, the subresource integrity standard allows to specify one or more hashes for \texttt{<script>} and \texttt{<link>} tags with an additional \texttt{integrity} attribute~\cite{W3C16b}. When a client's browser retrieves the content referenced with these tags and a correctly specified \texttt{integrity} is present, it refuses to process the content in case the predefined hashes do not match the hash of the content retrieved. Furthermore, the \textit{content security policy} (CSP) dialect can be used to instruct browsers to enforce the subresource integrity constraint for all external (non-inline) scripts \cite{Mozilla18c}. In this case a browser refuses to execute a JavaScript either in case a valid \texttt{integrity} attribute is missing or the integrity check fails. Additional information can be passed to browsers by using the \texttt{crossorigin} attribute from the CORS standard to tell browsers that credentials (including cookies and certificates) are not transmitted through the tags equipped with the subresource integrity checks. By using the symbol $\hash$ to again denote a message digest, $g(\cdot)$ a base64-encoding function, and referring to the earlier example in \eqref{eq: url}, the following excerpt can be used to illustrate the syntax of a subresource integrity check enforced for a cross-origin JavaScript:
\begin{align}
&\texttt{<script src="https://example.com/javascript.js"}
\\ \notag
&~~\texttt{integrity="sha256-}g(\mathcal{H})\texttt{"}~\texttt{crossorigin="anonymous"></script>}
\end{align}

Thus, the basic idea is simple but not bulletproof. The standard mentions three potential weaknesses \cite{W3C16b}. The first weakness is cryptographic: potential hash collisions undermine the foundations of all integrity checks done with a particular algorithm. The second weakness relates to transmissions: a malicious proxy can obviously strip the attributes in case plain HTTP is used or the context is otherwise insecure (see \cite{Conti16} for a survey of these man-in-the-middle scenarios). The third weakness originates from information leakages: it may be possible to deduce about the critical parts of a website protected by integrity checks. By repeatedly loading resources for which integrity checks are enforced, it may be possible to gain information about whether the content protected is static or dynamic. Particularly in case CORS is not used in conjunction with the integrity checks, these information leakages may allow an attacker to eventually guess authentication details \cite{W3C16b}, for instance. In addition to these three explicitly mentioned weaknesses, so-called browser cache poisoning may potentially circumvent the integrity checks~\cite{JiaChen15}. As always, there may be also other already known or yet unknown weaknesses affecting the subresource integrity standard.

\subsection{Practical Integrity Challenges}

There are many practical challenges for widespread integrity checking of cross-origin scripts. Arguably, the cardinal challenge has never been the lack of technical solutions and standards, but rather the adoption of these solutions and standards among clients, servers, software producers, web developers, and numerous other actors involved. In terms of the standardized solutions, a major practical challenge relates particularly to web development practices and the manual work entailed in the implementation and enforcement of the solutions~\cite{Prokhorenko16}. The pocket-sized analytical framework from the previous section can be used to exemplify three practical scenarios on how integrity may vary for cross-origin scripts.

First, two distinct websites $\WebPage\not\equiv\WebPage^\prime$ may include a script with the same unique $\URL$ pointing to the same unique $\JavaScript$ with the same unique $\hash$. Second, it is possible that two unique URLs from two distinct websites point to the same unique JavaScript content, possibly at different times $t$ and $t + k$. In this case
\begin{equation}\label{eq: two urls same hash}
h(f(\URL, t)) 
\equiv h(f(\URL^\prime, t + k)) 
\equiv\hash_t
\end{equation}
holds for $\URL\in\WebPage$, $\URL^\prime\in\WebPage^\prime$, $\URL\not\equiv\URL^\prime$, and $\WebPage\not\equiv\WebPage^\prime$. These cases occur because web developers may copy JavaScripts from different Internet sources to their own websites. Furthermore, it is relatively common that a same small script is used in various parts of a website, such that \eqref{eq: two urls same hash} holds for 
$\URL\in\WebPage$, $\URL^\prime\in\WebPage$, and $\URL\not\equiv\URL^\prime$. 

In other words, certain JavaScripts are included by many websites, scripts from one website may be plagiarized to other sites, and some JavaScripts are used in multiple parts of a single website. All three scenarios have security implications. Besides outright duplicates, the common occurrence of approximately highly similar JavaScript code (a.k.a. code clones)~\cite{CheungRyu16} is problematic because a cloned script may contain vulnerabilities. Cloned scripts are also unlikely to be rigorously maintained already due to the lack of strict references to the original sources for which vulnerabilities may be fixed by the original authors. The problem is not only theoretical: recent Internet measurement studies indicate that many websites include vulnerable and outdated JavaScript libraries \cite{Lauinger17}. Regardless whether a vulnerable script is cloned or original, the potential attack vector also increases in case the script is used in multiple parts of a website. 

The inclusion of certain cross-origin scripts by numerous websites raises the attack surface to an entirely different level. By compromising a popular CDN used to distribute JavaScript code, arbitrary code may be injected to thousands (or even millions) of websites, and this code may be executed by millions (or even billions) of clients. Even though this scenario has not fortunately realized, the theoretical possibility cannot be ruled out. On the other hand, the scenario has already realized on the side of privacy: ``arbitrary code'' is executed by billions of web clients due to the inclusion of cross-origin scripts by millions of websites. By assumption, it is also the third-party tracking infrastructures and web advertisements served via these infrastructures that make it difficult for web developers to enforce temporal integrity checks. In fact, it has been argued that web developers no longer even know who they are trusting with their remote JavaScript inclusions \cite{Kumar17, Ruohonen18PST}. Given this motivation, the remainder of this paper focuses on the question of how common temporal integrity changes are in reality. 

\section{Data}\label{section: data}

In what follows, the dataset is elaborated by discussing the sampling and polling routines used to assemble the dataset and the measurement framework for it.

\subsection{Sampling}

By following common research approaches for retrieving JavaScript code \cite{Krueger12, Nikiforakis12}, the initial collection of JavaScripts was done by sampling ten thousand unique second-level domain names from a ranking list made available by Cisco~\cite{Cisco18}. It is worth remarking that Cisco's lists have been used also previously~\cite{Mayer17} as an alternative to Alexa's lists, which are no longer available free of charge. More importantly, each domain in the list was transformed to a second-level domain name. This transformation is justified because the ranks are based on the volume of \textit{domain name system} (DNS) traffic passing through Cisco's (OpenDNS) servers. For this reason, the list contains separate entries for example for \texttt{microsoft.com} and its subdomains such as \texttt{data.microsoft.com} and \texttt{ipv6.microsoft.com}. 

Five further remarks are required about the sampling. To begin with, (a) each domain was queried with the \texttt{http} scheme. Thus, Microsoft's main domain was queried by passing \texttt{http://microsoft.com} to a browser, for instance. That said, (b) it is important to emphasize that redirections were followed for all queries. In terms of the running example, the URL requested was actually redirected to a location \texttt{https://www.microsoft.com/fi-fi/}. These redirections involved both the DNS and the HTTP protocol; typically, HTTP redirections upgraded the \texttt{http} scheme requested to HTTPS connections, while either HTTP or DNS redirections occurred to the subdomains (such as the \texttt{www}-prefixed ones) of the requested second-level domains. The accounting of these redirections is essential for deducing about cross-origin scripts. In addition: when dealing with mostly dynamic content in the contemporary Web, a JavaScript-capable browser is required particularly for executing inline JavaScripts, which may interfere with the execution of external scripts \cite{Nikiforakis12}. Therefore, (c) a custom WebKit/Qt-powered headless browser was used with JavaScript enabled for all queries. By again following common practices~\cite{Ruohonen17EISIC, Ruohonen18PST}, (d) a $30$ second timeout was used for all queries to ensure that the majority of scripts were successfully executed. Finally, (e) all domains were queried two times in order to account temporary failures.

\subsection{Polling}

The domains sampled were used to construct a pool for temporal integrity polling. To construct the pool, all scripts were collected from each domain sampled, but only cross-origin scripts were qualified to the polling pool. All cross-origin comparisons were done based on the visited URLs (and not the requested ones), which were used also for transforming relative URLs to absolute ones. Thus, an initial redirection to a HTTPS connection or a subdomain did not qualify an entry to the pool. During the construction of the pool, each cross-origin $\URL$ in $\textmd{\texttt{<script src="}}\URL\textmd{\texttt{">}}$ was downloaded via a GET request. After observing that many prior test requests failed when the \texttt{query} and related fields were stripped, all downloads were made with the exact same URLs used in the websites sampled. Although no extensive attempts were made to verify that a given URL actually pointed to a valid JavaScript, \texttt{Content-Type} and related HTTP header fields were recorded for the initial downloads. In addition, each download was passed through a program for checking the \textit{multipurpose Internet mail extension} (MIME) type. Finally, the downloads that returned non-empty buffers with a HTTP status code $200$ were used to construct the polling pool, $\left[ \URL_1, \ldots, \URL_{35417} \right]$. The URLs within the pool were then polled with HTTP GET requests consecutively for $T = 10$ days starting from March 23, 2018.

The empirical focus is on the following representation of the polling pool:
\begin{equation}
\left[ 
h(f(\URL, t)), 
h(f(\URL, t + 1)), 
\ldots,
h(f(\URL, t + T - 1)) 
\right]_i , 
\quad i \in [ 1, 35417 ] ,
\end{equation}
where $\URL$ satisfies \eqref{eq: cross-origin URLs} with respects to a sampled $\WebPage$. To account for temporary transmission errors, domain name resolution failures, and other related networking shortages, the cases marked with a symbol $\NA$ in \eqref{eq: download} were first removed from each vector. After this removal,  each vector was transformed to a set, such that only unique hashes are observed for each script. (None of the transformations resulted a $\emptyset$, which would indicate that all polls would have failed.) This simple operationalization provides a straightforward way to observe the temporal integrity of cross-origin scripts: if $\hashset$ denotes a set of unique hashes, the temporal integrity of a given script residing at $\URL$ was intact during the polling period if and only if $\vert\hashset\vert = 1$. Although the polling period allows to only observe a rather short time span, a simple subtraction $\vert\hashset\vert - 1$ gives the number of changes.

\section{Results}\label{section: results}

In what follows, the main empirical insights are summarized by presenting a few descriptive statistics on the dataset and then discussing the classification results.

\subsection{Descriptive Statistics}

\begin{figure}[th!b]
\centering
\includegraphics[width=\linewidth, height=4cm]{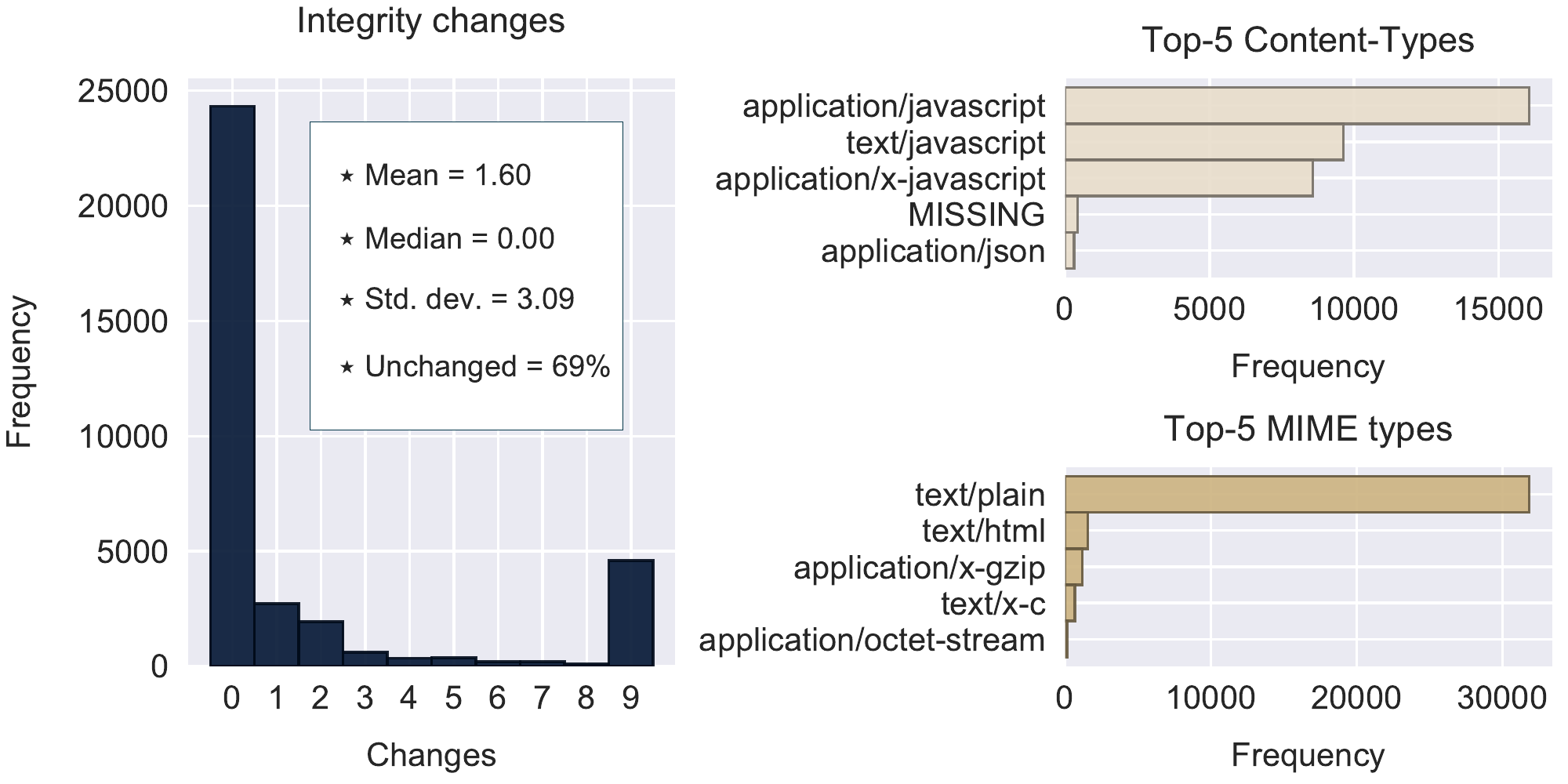}
\caption{Temporal Integrity Changes ($\vert\hashset\vert -1$) and Buffer Types}
\label{fig: changes}
\end{figure}

Temporal integrity changes are relatively common: more than a quarter of the polls indicated at least one integrity change. The shape of the distribution in Fig.~\ref{fig: changes} is also interesting: it seems that integrity changes may tend to converge toward a bimodal distribution. In other words, the dataset contains a majority class of scripts for which temporal integrity remained intact, and a minority class for which each daily download resulted in a different hash. The two right-hand side plots also tell that the contents downloaded are indeed mostly JavaScripts.

\begin{figure}[th!b]
\centering
\includegraphics[width=\linewidth, height=4.5cm]{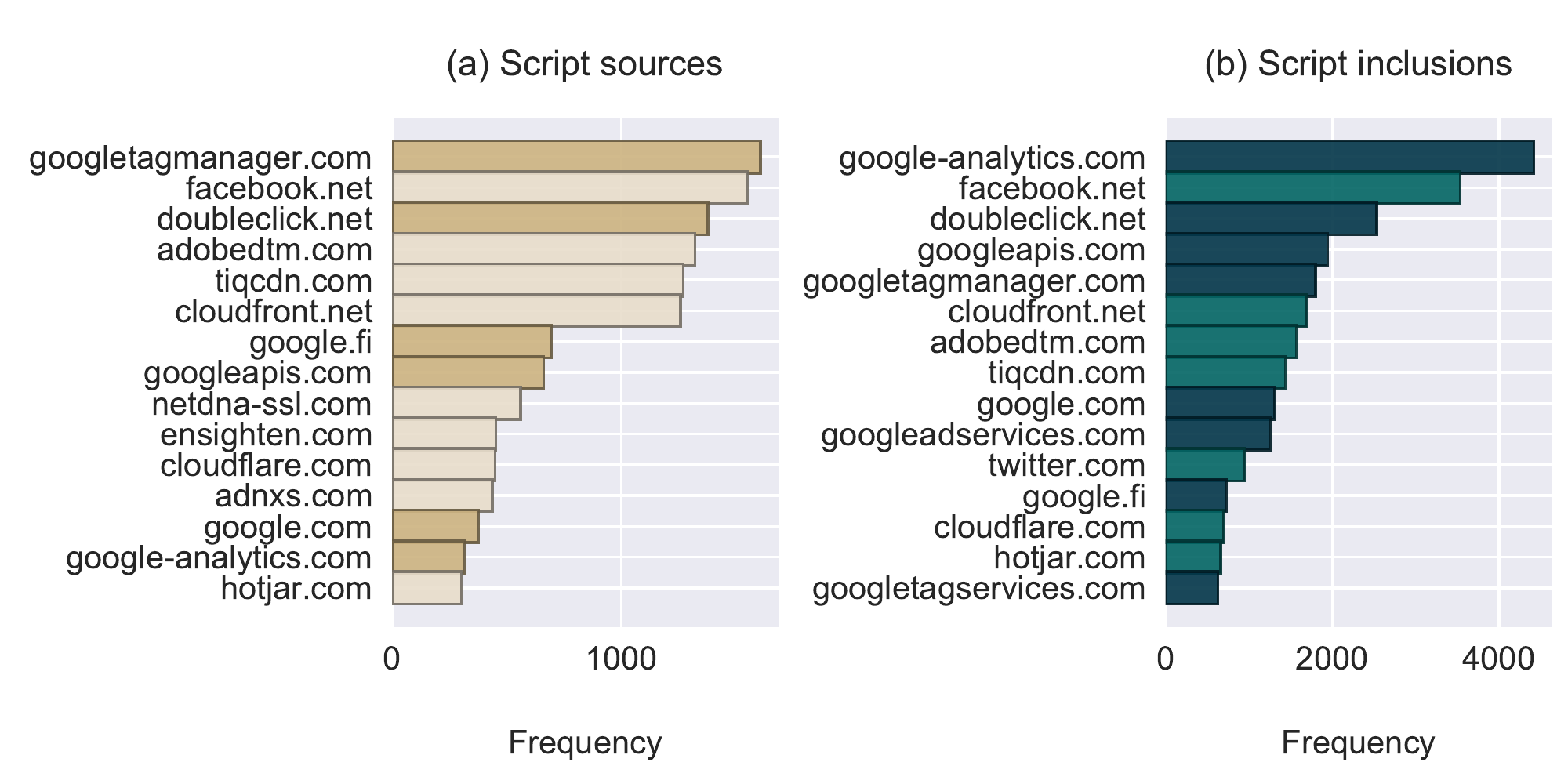}
\caption{Top-15 Second-Level Domains (ranks based on (a) the aggregation from the \texttt{host} fields of the URLs polled and (b) the number of times the aggregated \texttt{host} fields were included by the cross-origin \texttt{<script>} tags of the second-level domains sampled)}
\label{fig: domains}
\end{figure}

The temporal integrity changes observed go hand in hand with the lack of integrity checks. There are one-to-many references from the URLs polled to the \texttt{<script>} tags that included the sources behind the URLs with cross-origin references. To briefly probe the attributes within these tags, the percentage share of URLs with at least one ``back-reference'' to a given attribute can be used. Given the few interesting, JavaScript-specific attributes, the shares are: $42.2\%$ for \texttt{async}, $3.7\%$ for \texttt{defer}, $2.4\%$ for \texttt{crossorigin}, and $0.34\%$ for \texttt{integrity}. Thus, it is safe to generalize that subresource integrity checks are rarely used in the current Web. Given the integrity changes observed, widespread future adoption of the subresource integrity standard seems also somewhat unlikely.

However, this tentative prediction partially depends on the cloud service and CDN companies who are hosting and distributing popular JavaScripts. For pointing out the main players, Fig.~\ref{fig: domains} shows the second-level domain names of the fifteen most frequent sources behind the cross-origin scripts observed. The two plots largely confirm the existing wisdom: common locations of remote JavaScript code are extremely concentrated, tracing only to a few companies~\text{\cite{Kumar17, Nikiforakis12}}. In particular, Google continues to be the leading distributor of common JavaScript snippets, although Facebook has recently been catching up.

\subsection{Classification}

\begin{figure}[th!b]
\centering
\includegraphics[width=\linewidth, height=6.1cm]{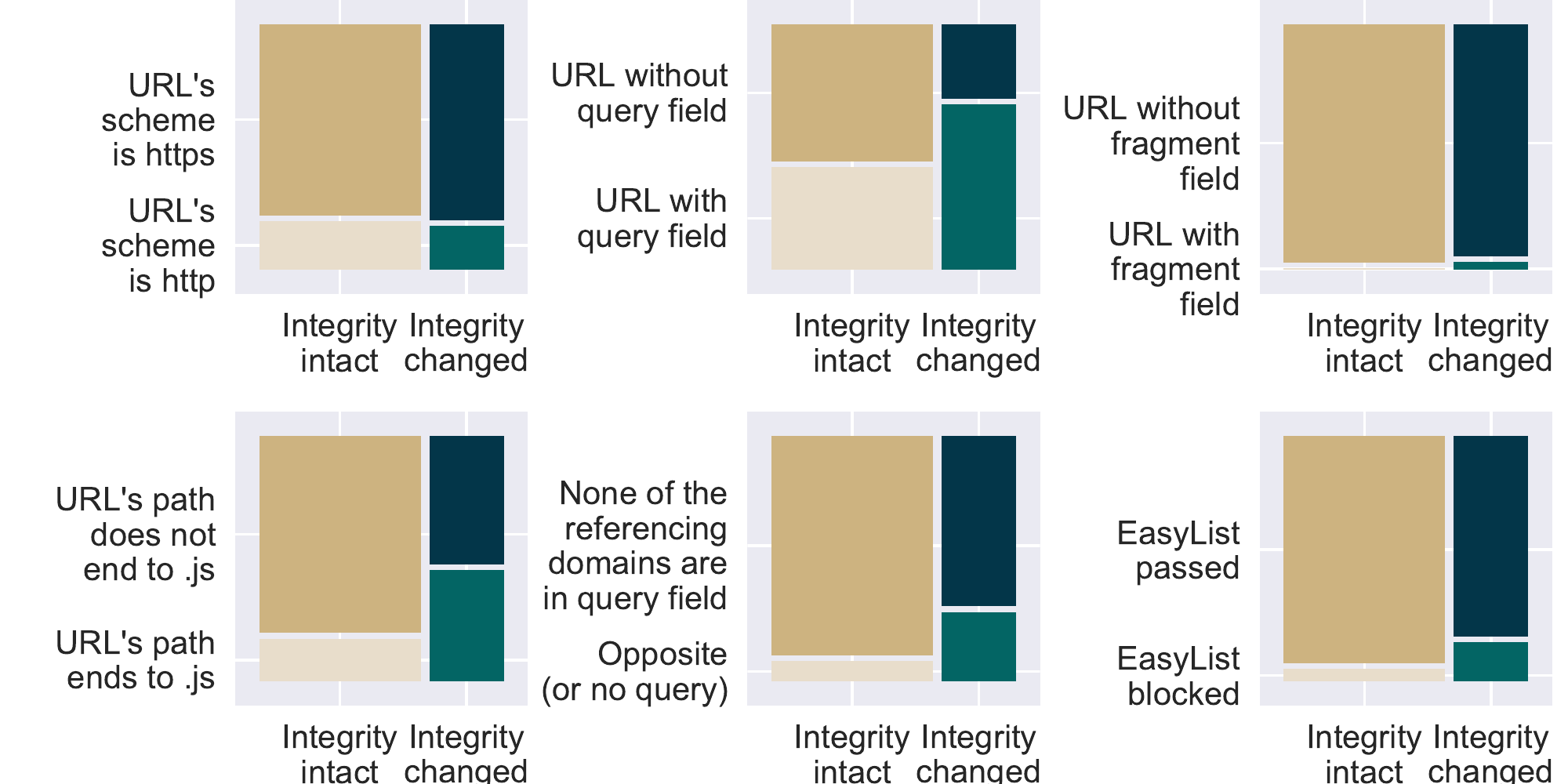}
\caption{Temporal Integrity According to a Few Dichotomous Metrics (see main text)}
\label{fig: mosaics}
\end{figure}

It is interesting to examine how systematic the temporal integrity changes are statistically. For this purpose, the conditional probability that a change occurs, $\vert\hashset\vert > 1$, is a sensible measurement target. As for features potentially explaining a change, a good point of reference is provided by the literature on classifying URLs pointing to malware and phishing websites. This literature typically operates with numerous simple metrics extracted from URLs, DNS, and related sources~\cite{Abdelhamid15, Ma09b, Vasek13}. To illustrate a few of such metrics, Fig.~\ref{fig: mosaics} displays six so-called mosaic plots. Interpretation of the plots is easy: in each plot the area of a rectangle corresponds with the frequency of a cell in a contingency table.

\begin{table}[th!b]
\centering
\caption{Metrics ($\dichotomous$ for dichotomous and $\continuous$ for continuous scale, $f(x) = \log(x + 1)$ applied for all metrics with $\continuous$ scale, no additional scaling or centering for classification)}
\label{tab: metrics}
\begin{small}
\begin{tabular}{lcl}
\toprule
Metric & Scale & Description and operationalization \\
\hline
INCL & $\continuous$ & Number of sampled domains that included a script with a $\URL$. \\
SLEN & $\continuous$ & Character count of the buffer during the first download. \\
BLCK & $\dichotomous$ & True if a $\URL$ would be blocked by a common ad-blocking list \cite{EasyList18}. \\
QURL & $\dichotomous$ & True if a \texttt{query} field is present in a $\URL$. \\
QDOM & $\dichotomous$ & True if any of the domains including a script's $\URL$ appears in \texttt{query}. \\
NOJS & $\dichotomous$ & True if a \texttt{path} field of a $\URL$ does \textit{not} end to a \texttt{.js} character string. \\
ULEN & $\continuous$ & Character count of a whole $\URL$ used for the polling. \\
UNUM & $\continuous$ & Number of numbers $(0, \ldots, 9)$ appearing in a whole $\URL$. \\
DNUM & $\continuous$ & Number of domains in a \texttt{host} field (excl.~the top-level domain and IPv4s). \\
DTOP & $\dichotomous$ & A dummy variable for each of the domains in the plot (a) in Fig.~\ref{fig: domains}. \\
\bottomrule
\end{tabular}
\end{small}
\end{table}

\begin{figure}[th!b]
\centering
\includegraphics[width=\linewidth, height=3.0cm]{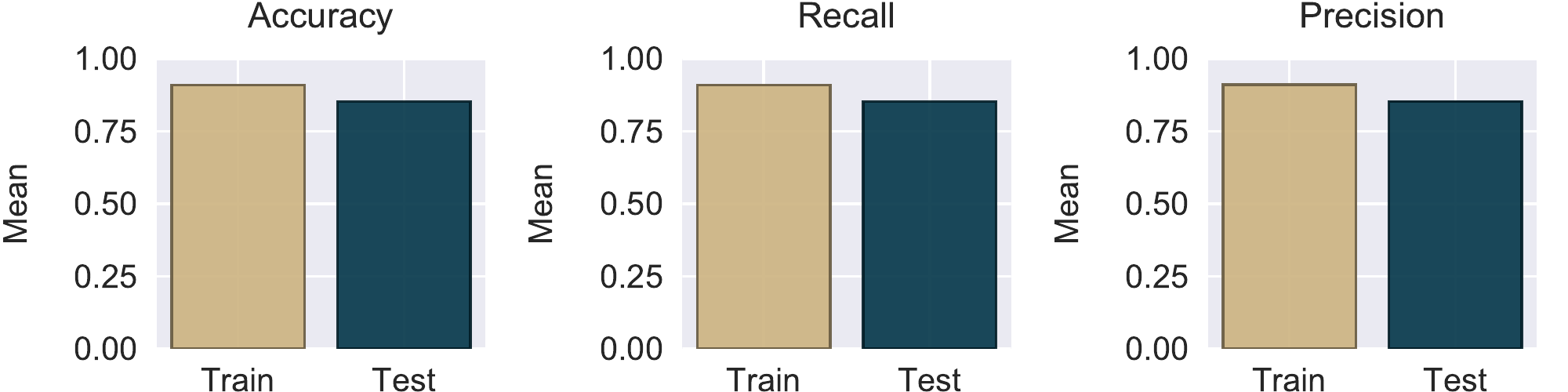}
\caption{Classification Performance ($35417$ script download URLs, $24$ metrics, decision tree classifier~\cite{scikitlearn}, maximum tree depth restricted to $15$, and $10$-fold cross-validation for each of the $100$ random samples with under-sampling from the majority class)}
\label{fig: classification}
\end{figure}

Although the first plot indicates that HTTPS does not explain temporal integrity changes, it is still noteworthy---and troublesome---that many of the scripts were included by the websites sampled with the \texttt{http} scheme. As was discussed in Subsection~\ref{subsec: integrity of cross-origin javascripts}, subduing the use of plain HTTP is a prerequisite for sound integrity checks. The two remaining plots on the upper row foretell about an association between integrity changes and the presence of \texttt{query} fields but not \texttt{fragment} fields. These observations are reinforced by the three plots on the second row. In particular, a \texttt{path} ending to \texttt{.js} is associated with integrity changes, the domains sampled often appear in the \texttt{query} fields of those URLs whose content changed during the polling, and, finally, the probability of a temporal integrity change is slightly higher for URLs blocked by a common ad-blocking list \cite{EasyList18} according to an offline parser \cite{adblockparser18}. These observations hint that temporal integrity changes are typical with respect to scripts used for advertisement and tracking purposes. While this conclusion may seem unsurprising, it is important in terms of future adoption of integrity checks for cross-origin JavaScript code. Because many websites rely on advertisements and analytics for business reasons, but the corresponding scripts tend to violate temporal integrity premises, it seems that many websites are simply unable to enforce subresource integrity checks---even when these would be widely endorsed by web developers.

After empirically reviewing over $25$ metrics, the ten metrics enumerated in Table~\ref{tab: metrics} turned out to be relevant for statistical prediction. The metrics that did not improve prediction include all of the standardized \texttt{<script>} attributes, all of the \texttt{Content-Type} and MIME types present in the sample, top-level domain names extracted from the URLs, and numerous dummy variables such as whether a \texttt{host} field refers to an Internet protocol (IPv4) address. Given the limited amount of information used for predicting whether a temporal integrity change occurs, the results summarized in Fig.~\ref{fig: classification} are even surprisingly good. The average classification accuracy is $0.85$. It can be concluded that the temporal integrity of cross-origin JavaScripts vary systematically, and that it is possible to predict whether a change occurs to a reasonable degree even with limited information.

\section{Discussion}\label{section: discussion}

This paper presented the first empirical study on temporal integrity of remote, cross-origin JavaScript code commonly used in the current Web. According to the empirical results, temporal integrity changes---or, depending on the viewpoint, temporal integrity violations---are relatively common. Given over $35$ thousand URLs observed in a short polling period of ten days, about 31\% of the JavaScript content behind the URLs witnessed at least one temporal integrity change. One way to digest this result is to simply state that arbitrary code is commonly executed on the client-side of the current Web. Because temporal integrity is not guaranteed, a cryptomining script \cite{Eskandari18}, for instance, can easily replace an existing legitimate script without any alerts for the clients executing the script.

There are many potential solutions but all of these contain limitations. The simplest solution would be to block all cross-origin content on the client-side, but this would severely impact functionality and user experience. Another solution would be to transform cross-origin \texttt{<script>} tags to \texttt{<iframe>} tags~\cite{Bugliesi17}, but this solution has performance implications, and it cannot solve the privacy problems. Analogously: using code clones solves the reliance on dynamically loaded third-party code, but at the expense of maintenance and the security risks entailed by in-house maintenance of third-party code \cite{Nikiforakis12, SomeBielova17}. The subresource integrity standard offers a further option. As was discussed and empirically demonstrated, widespread adoption of the standard faces many practical obstacles, however. One obstacle affecting the standard---as well as this paper---is the lack of context behind the temporal integrity changes. In other words, a different hash will result upon fixing a vulnerability in a third-party JavaScript library or making a cosmetic change to such a library. Deducing about the nature of temporal changes would be a good topic for further research, although the commonplace obfuscation of JavaScript code makes the topic challenging to say the least.

To put technical details aside, it might be also possible to refine the underlying ideas presented in the standard. The standard leaves the enforcement of integrity checks to the server-side, but there are no theoretical reasons why clients could not enforce the checks themselves based on a trusted collection of scripts. After all, code signing has a long history elsewhere in the software industry~\cite{Catuogno15}. Given that both web developers and the JavaScript library ecosystem are still taking their first steps toward systematic dependency management and rigorous vulnerability tracking \cite{Lauinger17}, code signing seems like a good long-term goal rather than an immediately applicable solution, however. But for large CDNs and companies such as Google and Facebook, signing the JavaScript code included by hundreds of millions of websites might be possible even today. Another question is whether temporal integrity is in the interests of these companies---if clients would no longer blindly execute arbitrary code, user tracking would be more difficult. In this sense, there exists a classical trade-off between security and privacy,  but the current balance that violates privacy undermines also security.

\bibliographystyle{splncs03}

\end{document}